# End-to-End Available Bandwidth Measurement Tools: A Comparative Evaluation of Performances

Ahmed Ait Ali, Fabien Michaut, Francis Lepage

CRAN (Centre de Recherche en Automatique de Nancy)
UMR-CNRS 7039
Faculté des Sciences et Techniques - BP 239
54506 Vandoeuvre Cedex
`{ahmed.aitali, fabien.michaut,`
`francis.lepage}@cran.uhp-nancy.fr`

**Abstract.** In recent years, there has been a strong interest in measuring the available bandwidth of network paths. Several methods and techniques have been proposed and various measurement tools have been developed and evaluated. However, there have been few comparative studies with regards to the actual performance of these tools. This paper presents a study of available bandwidth measurement techniques and undertakes a comparative analysis in terms of accuracy, intrusiveness and response time of active probing tools. Finally, measurement errors and the uncertainty of the tools are analysed and overall conclusions made.

## 1  Introduction

In data communication networks, high available bandwidth is useful because it supports high volume data transfers, short latencies and high rates of successfully established connections. Obtaining an accurate measurement of this metric can be crucial to effective deployment of QoS services in a network and can greatly enhance different network applications and technologies.

Several applications need to know the bandwidth characteristics of their network paths. For example, some peer-to-peer applications need to consider available bandwidth before allowing peers to join the network. Overlay networks can configure their routing table based on the available bandwidth of the overlay links. Network providers lease links to customers and the charge is usually based on the available bandwidth that is provided. Service Level Agreements (SLAs) between providers and customers often define service in terms of available bandwidth at network boundaries. Available bandwidth is also a key concept in congestion avoidance algorithms and intelligent routing systems.

Techniques for estimating available bandwidth fall into two broad categories - passive and active measurement. Passive measurement is performed by observing existing traffic without perturbing the network. It needs to process the full load on the link and requires access to all intermediary nodes in the network path to extract end-to-end information. Active measurement on the other hand, directly probes network proprieties by generating the traffic needed to make the measurement. Despite the fact that active techniques inject additional traffic on the network path; it is more suitable to use active probing measurement in order to measure end-to-end available bandwidth.

Many different active probing techniques and tools for available bandwidth measurement have been developed and evaluated. However, a sufficient number of studies comparing the performance of these tools have not been carried out.

In this paper, a study of available bandwidth measurement methodologies is presented and a comparative analysis in terms of accuracy, intrusiveness and response time of active probing tools for this metric is achieved. In comparison with previous work, this paper analyses tool measurement uncertainties, investigates the sources of observed errors that are likely to be inherent in delay measurement and presents some probe pattern weaknesses that can affect measurement accuracy. We have collected our measurements using a simple testbed configuration that allows us to test measurement tools (Spruce, Pathload, IGI, Pathchirp) with the same parameters.

The remainder of the paper is structured as follows. Section 2 presents the most prevalent bandwidth-based metrics. Measurement techniques and tools are described in section 3. Section 4 presents the results of the experiments for a performance evaluation of the studied tools. Measurement uncertainties and the sources of the observed errors are analysed in section 5. Finally, we conclude in section 6.

## 2 Bandwidth-Based Metrics

In practice, four bandwidth measurements can be performed, namely the capacity/raw bandwidth of a link, the end-to-end capacity of a path, the available bandwidth of a link and the available bandwidth of a path. In the following section we will define available bandwidth parameters and present the techniques and tools for this metric.

$P$ is a network path from source $S$ to destination $D$. $P$ is a sequence of $N$ store-and-forward links $Ll_1, Ll_2, \ldots Ll_N$. We assume that $P$ is fixed and unique (no routing changes or multipath forwarding occur during the measurement).

− The capacity of the link, denoted $C_i$, is the maximum possible IP layer transfer rate at that link. The end-to-end capacity of the path is then the maximum IP layer rate that this path can transfer from the source $S$ to the sink $D$:

$$C = \min C_i \qquad (1)$$

− The available bandwidth of a link defines the unused capacity of this link during a certain time period. We assume that link $i$ is transmitting $C_i u_i$ bits during a time interval $T$. $u_i$ is the utilisation rate of this link during $T$, with $0 \leq u_i \leq 1$. The available bandwidth $A_i$ of the link $i$ is:

$$A_i = C_i(1-u_i) \qquad (2)$$

− The available bandwidth $A$ of the path $P$ during the time interval $T$ is the minimum of the available bandwidth of all links that comprise $P$:

$$A = \min_{i=1...N}\{C_i(1-u_i)\} = \min_{i=1...N} A_i \qquad (3)$$

## 3 Related Work

To assess and monitor network bandwidth metrics, many software tools have been developed. They are based on different principles and which integrate various techniques. The main purpose of this section is to describe the most commonly used techniques and methodologies and provide some examples of tools using them. This section also aims to present previous work done in projects comparing measurement tools.

### 3.1 Measurement Techniques and Methodologies

Available bandwidth estimation techniques can be divided into two categories: self-induced congestion based techniques and cross traffic estimation based techniques.

Self-induced congestion based techniques assume FIFO queuing at all routers along the path, cross traffic follows a fluid model and average rates of cross traffic change slowly. If a source sends probes to a destination at a rate $R$ less than $A$, probes will experience similar delays. On the other hand, if $R$ is greater than $A$, probes will queue in the network and experience increasing delays. This technique is based on the observation that the delays of successive probing packets increase when the probing rate exceeds the available bandwidth in the path. It consists in probing the network at different rates and detecting (at the destination) the point at which delays start to increase. At this point, probing rate is equal to the available bandwidth. Pathload [1],[2] and PathChirp [5] implement this technique. In their work D.Katabi & al referred to Probe Rate Model rather than self-induced congestion [3].

Pathload introduces a technique based on Self Loading Periodic Stream (SLoPS) [4]. The algorithm consists in sending a stream of packets to the receiver. The receiver then measures the delay of each received packet and analyses its variation. If the delay is constant, an other stream is sent to the receiver at a greater rate. If the delay increases, the next stream is then sent to the receiver at a rate between the two previous values. This technique is repeated and the algorithm converges by dichotomy to the available bandwidth value. PathChirp proposes sending an exponentially spaced 'chirp' probing train. The main advantage of this approach is to minimize the probing traffic load. Indeed, a single chirp is able to probe the network at different rates.

Cross traffic estimation based techniques assume that the capacity $C$ of the path is known and the bottleneck is both the narrow and the tight link (respectively the link with the smallest capacity and the link with the smallest available bandwidth). These techniques and tools are based on the Probe Gap model (PGM) [3] that consists in capturing the relationship between the dispersion of a packet-pair and the cross traffic rate $C_T$ at the bottleneck link of a path [6]. They begin by estimating the cross traffic at the bottleneck and then compute the available bandwidth as the difference between the path capacity and the cross traffic rate: $A=C-C_T$.

These techniques are implemented in IGI [7] and Spruce [3]. IGI uses packet trains i.e. a longer sequence of evenly spaced packets to probe the network. Experiments carried out in [7] have shown that the optimal initial gap $D_{in}$ is obtained when the average output gap $D_{out}$ equals the initial gap. IGI starts by sending packet-pairs with a small $\Delta_{in}$ and increases it until the average output gap equals the initial gap. The available bandwidth is estimated using formula (4).

$$A = C - C_T = \frac{C(D_{in} - D_{out}) + L}{D_{in}} \qquad (4)$$

To cope with packet-pairs dependence, Spruce sends a Poisson process of packet-pairs. Additionally, spruce adjusts the average inter-pair gap to ensure that the probe rate is a minimum of 240 Kbs and 5% of the end-to-end capacity. Spruce uses formula (5) to estimate the available bandwidth.

$$A = C - C_T = \left(2 - \frac{D_{out}}{D_{in}}\right).C \qquad (5)$$

### 3.2 Comparison Projects

In available bandwidth measurement some performance comparison studies have been proposed. For example, Strauss & al [3] compared Spruce to IGI and Pathload. The comparison focused on accuracy, failure patterns and probe overhead. They used MRTG data to perform the comparison over 400 different Internet paths. The authors concluded that Spruce is more accurate than Pathload and IGI. Their measurement data showed that Pathload tends to overestimate the available bandwidth whereas IGI becomes insensitive when the bottleneck utilisation is high.

Shrarem & al [10] evaluated publicly available tools of available bandwidth measurement on high-speed links. They compared Abing, Spruce, Pathload and Pathchirp in an isolated high-speed testbed developed by CAIDA researchers in collaboration with the calNGI Network Performance Reference Lab [11] and found that Pathload and Pathchirp are the most accurate tools under the conditions of their experiments.

Hu and Steenkiste [7] developed two available bandwidth measurement tools, IGI and PTR, and tested them on 13 Internet 100 Mbs paths. They compared them to pathload performance using Iperf [12] as a reference tool. Because Iperf was not absolutely accurate, there is still great uncertainty as to the performance of IGI and PTR.

In this paper, the performances of Spruce, Pathload, Pathchirp and IGI are compared in an isolated testbed configuration. Compared to previous work, what is original about this paper is that we study uncertainties of obtained measurements and explain some error sources that are mainly inherent to pattern failures and to packets time stamping operations.

## 4  Comparative Evaluation

This section aims to compare the performance of the measurement tools described above regarding accuracy, response time and intrusiveness. For that purpose, we ran several simple experiments on an isolated testbed configuration. The topology is shown in figure1 where $P_s$ and $P_d$ are the probing source and destination.

### 4.1  Methodology and Measurement Testbed

To compare the performance of the measurement tools studied above, a set of experiments are carried out on an isolated network that comprises three Cisco 1700 series routers connected through FastEthernet links. The network built for tests, comprises sender and receiver hosts, belonging to different LANs on which several cross-traffic sources are active. The used hosts are equipped with Debian Linux (kernel version 2.4.25) and have the same hardware configuration. A couple of hosts are used as the sender and the receiver part of the measurement tool, while the other hosts are used as the source and the destination for the cross-traffic. The choice of an isolated network is motivated by the need to totally control the network under test and to respect the initial assumptions made by the authors of the different measurement tools.

End-to-end available bandwidth measurement tools in these experiments consist of separate user-level sender and receiver parts. The sender part is set up on $P_s$ and the receiver part on $P_d$. $C_s$ and $C_d$ are used to generate cross traffic using an MGEN traffic generator. The traffic analyser Ethereal is installed both on tool and traffic generator receiver parts. It is used to report measurement traffic load and response times of each tool and to verify the cross traffic throughput generated by the traffic generator. Each measurement tool has been run over more than 30 measurement sessions.

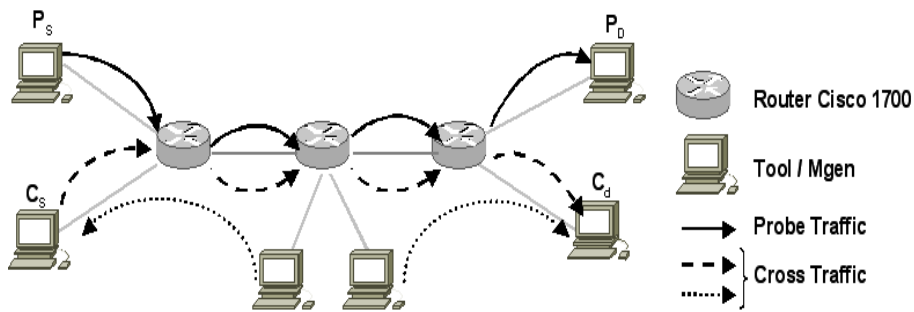

**Fig. 1.** Available Bandwidth Measurement Testbed. The path studied is constituted of three Cisco 1700 Series routers. The probe traffic is carried out between $P_s$ and $P_d$. The cross traffic occurs between $C_s$ and $C_d$.

The data presented in this section is collected using four different measurement tools: Spruce, Pathload, IGI and Pathchirp. In order to evaluate the accuracy and performances of these active probing tools, we use an MGEN traffic generator to create constant cross traffic at a given rate. By changing this rate, it is be possible to have a full range of available bandwidth values. Varying the cross traffic rate from 100 to 0 Mbs in a 100 Mbs path will make the available bandwidth vary from 0 to 100 Mbs.

The experiments are undertaken by increasing the cross traffic rate with 5 Mbs increments in each measurement session. A total of 30 experiments were run for each available bandwidth value. The measurements were collected and result averages for each tool are presented in figure 2.

In figure 2 the dashed lines represent the available bandwidth of the path ($P_s$, $P_d$) which we expect and also represents the IP layer available bandwidth that depends on IP packet size and takes into account the overhead of layer-2 encapsulation and framing.

### 4.2 Performance Comparison

The main observation from figure 2a is that Pathload is inaccurate. It provides unstable estimates over or underestimating the available bandwidth. However, it reacts properly to cross traffic variation and follows the available bandwidth global trend. Pathload stops the measurement prematurely when available bandwidth is less than 10 Mbs (about 10% of the link capacity). This can be explained by the fact that Pathload integrates a packet loss detection mechanism that aborts the probe stream measurement when it encounters losses that exceed 10%. If this behaviour persists, Pathload stops the measurement and indicates to the user that the connection to the remote station is aborted. The same behaviour is encountered with experiments held in 10 Mbs paths. Pathload is unable to make measurements when available bandwidth is less than 1 Mbs.

Figure 2b plots the available bandwidth as measured by Pathchirp. It appears that Pathchirp reacts properly to the cross traffic variation but widely overestimates available bandwidth. Pathchirp provides good estimates of available bandwidth when the link utilisation is low and requires the specification of the response time parameter for each measurement session. Measurement tests using different response times have been performed.

Unlike the variation of the response time parameter, Pathchirp provides identical results. The measurement results presented in this paper using Pathchirp were obtained with 20 second time responses.

The results obtained for available bandwidth measurement using IGI are illustrated in figure 2c and show that IGI did not respond properly to cross traffic variation and, in most cases hugely underestimated available bandwidth.

Figure 2d illustrates the Spruce results for available bandwidth measurements and shows that Spruce closely tracks the available bandwidth and reacts reasonably well to cross traffic variation.

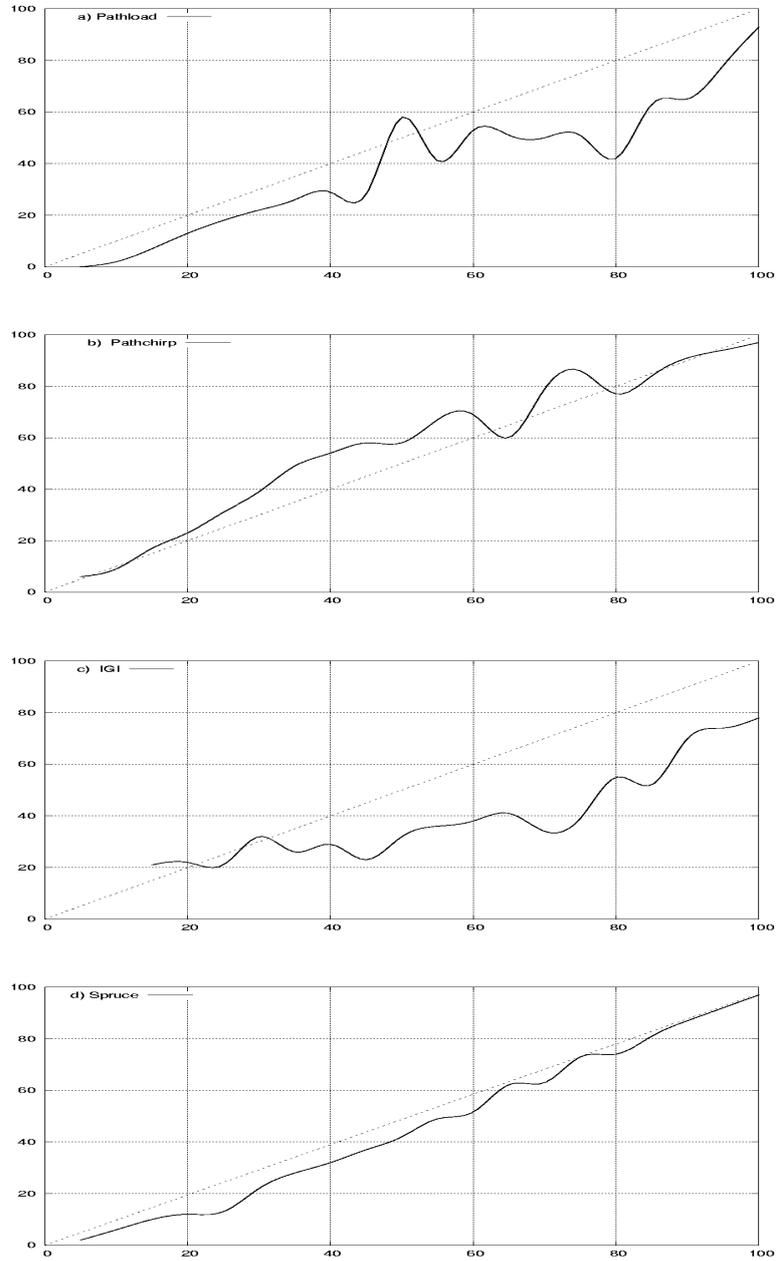

**Fig. 2.** Comparison of available bandwidth estimates using a) Pathload, b) Pathchirp, c) IGI and d) Spruce. The x-coordinate represents the expected available bandwidth value and y-coordinate is the measured available bandwidth reported by each tool.

Figure 3 shows the relative measurement errors of Spruce, Pathload, IGI and Pathchirp. We define the relative measurement errors as:

$$\varepsilon = \frac{|A_e - A_m|}{A_e} \qquad (6)$$

Where $A_e$ is the expected available bandwidth of the path and $A_m$ is the available bandwidth estimate.

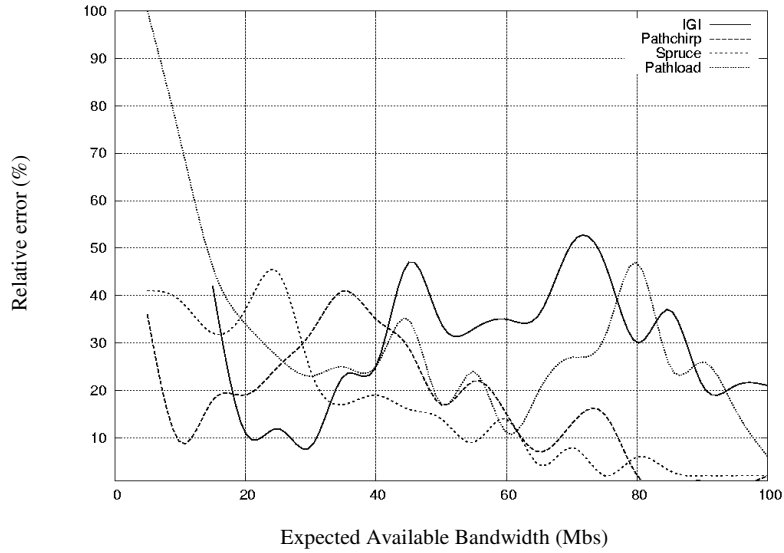

**Fig. 3.** Comparison of relative errors presented by Spruce, Pathload, IGI and Pathchirp. The x-coordinate represents the expected available bandwidth value and y-coordinate is the relative error reported by each tool.

Figure 3 shows that Pathload presents unstable relative measurement errors that change over a wide range of estimates. In some cases Pathload relative errors exceed 50%.

IGI and Pathchirp present unstable and high relative measurement errors. Almost all measurements have a relative error exceeding 40%. To measure the available bandwidth, IGI first measures the path capacity. Since the available bandwidth is calculated based on estimates of path capacity, errors on capacity measurement will lead to errors being produced on available bandwidth measurement. Thus, it is still difficult to draw conclusions about IGI measurement accuracy.

Pathchirp is more accurate than IGI, especially when the link utilisation is low. In some cases, Pathchirp measurement errors are less than 20%. Finally, figure 3 shows that almost 70% of Spruce measurements have a relative error under 20%. When the link utilisation is low relative measurement errors are less than 10%. Spruce appears to be the tool that reacts more appropriately to the available bandwidth variation and is the most accurate tool under our experimental conditions.

We considered several other parameters that may represent very important criteria when measuring the available bandwidth and which may potentially affect a user's decision when considering which tool to use. Measurement response time and tools intrusiveness are studied in the following. The trends are depicted in figure 4 and 5.

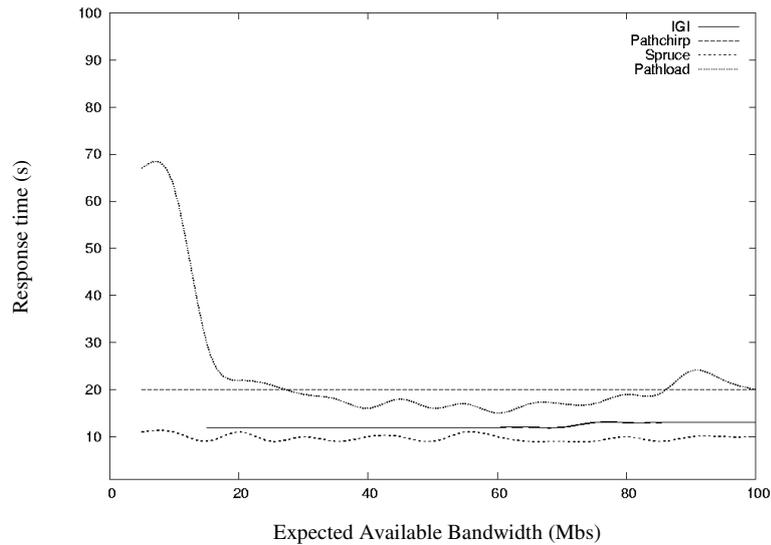

**Fig. 4.** Comparison of measurement response time using Spruce, Pathload, IGI and Pathchirp. The x-coordinate represents the expected available bandwidth value and y-coordinate is the measurement response time for each tool.

Available bandwidth is a parameter that varies over time. It is therefore essential to measure it as fast as possible. We define tool response time to be the average measurement time of all 30 measurement sessions. Figure 4 show that the observed measurement durations for Pathload are unstable and increase when the link utilisation exceeds 60%. Pathload's long measurement time is due to its convergence algorithm; it monitors changes on one-way delay of the probing streams and tries to converge its rate towards the available bandwidth value. The convergence slows down when probing packets experience different levels of congestion on the path.

The measurement duration experienced by the other tools is constant and independent of the link utilisation. IGI shows a 13 second response time. This measurement duration is given here on a purely informational basis since the response time of IGI includes path capacity measurement time. The response time of Pathchirp is an input parameter, it is specified by the user. Under the conditions of our experiments, the measurement time of Pathchirp is configured to 20 seconds. Experiments showed that increasing Pathload response times did not improve measurement accuracy. Figure 5 shows that Spruce is the fastest tool with only a measurement time of 10 seconds.

We define tool intrusiveness as the ratio of the average tool traffic rate to the path capacity. Figure 5 shows the measurement probe traffic load generated by each tool.

The main observation from this figure is that Pathload generates much more traffic than other tools and in some cases its measurement traffic can exceed 10% of the path capacity. Pathload intrusiveness is due to the SloPS algorithm that attempts to occupy all available bandwidth to extract path characteristics. Pathload measurement traffic increases when the link utilisation decreases. The average per-measurement probe traffic generated by IGI is 800 Kbs when the link utilisation is less than 40% and 200 Kbs in other cases. The results on IGI include the capacity measurement traffic. Therefore, the amount of measurement traffic generated by IGI in order to find the 'turning point' [7] and to calculate the available bandwidth is less than what which is plotted in figure 5. Spruce and Pathchirp generate constant and low amounts of measurement traffic that is around 200 and 300 Mbs, respectively.

In the next section, errors and uncertainties generally inherent to time stamping considerations were studied.

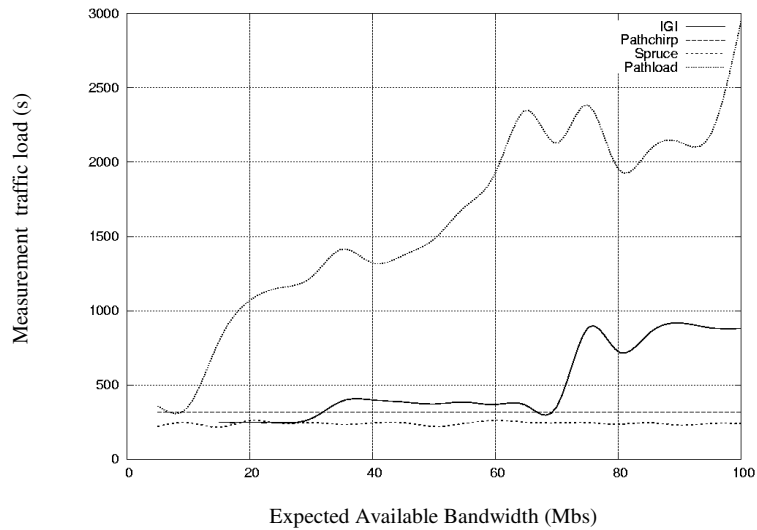

**Fig. 5.** Comparison of measurement traffic load generated by Spruce, Pathload, IGI and Pathchirp. The x-coordinate represents the expected available bandwidth value and y-coordinate is the amount measurement traffic generated by each tool.

## 5 Uncertainties and Errors Analysis

All the studied tools require careful scheduling of probe traffic. To be more precise, the initial inter-packet gap (in the case of cross traffic estimation based tools) and the probe stream rate (in the case of self-induced congestion based tools) must be accurate. To disregard these temporal constraints may result in measurement errors and make the degree of uncertainty associated with the measurement obtained higher would be expected. These errors and uncertainties depend mainly on the latency to timestamp the packet, move the packet from user to kernel space and transmit it on the network interface card. Furthermore, the OS scheduler may assign the computing resources to other concurrent processes between the timestamping and sending operations.

The timestamping operations are carried out by a system call "gettimeofday()". This function adds a supplementary latency to $\Delta t$ measurement. The estimated response time of "gettimeofday()" varies from 1 to 6 microseconds and is strongly dependent on hardware and software specifications (CPU frequency, OS version etc).
Spruce and IGI are more likely to be affected by the uncertainties depending on packet sending operations. Indeed, tools need to send packets with specific inter-packet gaps that are as small as a few hundred microseconds. A slight variation of $\Delta t$ of the order of 10 microseconds will produce a large relative error on $D_{in}$. The value of $D_{in}$ is used as a parameter in available bandwidth estimation. Thus, any error on $D_{in}$ will yield errors much larger than those typically expected in available bandwidth measurements.

We apply the differential calculus method in order to determine the uncertainties of the available bandwidth estimated by Spruce. Indeed, Spruce estimates available bandwidth using formula (5). Using the hypothesis that there is no error either on the path capacity estimation or on the final inter-packet gap measurement ($D_{out}$), the error $\Delta A$ on the available bandwidth is then:

$$\Delta A = \left| -\frac{C.D_{out}}{D_{in}^2} \right| \Delta D_{in} \qquad (8)$$

Assuming a path capacity of $C=97.5\ Mbs$, an initial gap error of $\Delta D_{in}=10\mu s$ and an available bandwidth of $A=50\ Mbs$ then we obtain an uncertainty $\Delta A=11.78Mbs$ on the available bandwidth measurement, corresponding to *23% of A*.

Pathchirp is based on the Probe Rate Model (PRM). Instantaneous rates involved by the probe packet chirps depend closely on the initial inter-packet gaps generated by Pathchirp.

By default, this measurement tool sends 1000 bytes packets to generate instantaneous rates varying from 10 to 200 Mbs. In order to do this, Pathchirp makes initial inter-packet gaps vary between 40 and 830 microseconds. Errors on initial gaps could lead to chirps being generated with instantaneous rates higher than expected. Since gaps are smaller and more sensitive to errors in low available bandwidth paths, the deterioration in the accuracy of initial gaps is more accentuated in this case. These errors in Pathchirp measurement delays will cause the available bandwidth paths to be overestimated.

Pathload uses one-way delay metrics to characterize the packet stream rates sent to the receiver part. The algorithm used in Pathload is based on the detection of an increasing trend in delays experienced by the packet streams going through the network path. Thus, errors in the rang of a few microseconds on packet timestamps have no effect on either the general behaviour of Pathload algorithm or on the final result reported by this tool.
Tools timestamp incoming packets at the receiver part in the same way as the packet-sending phase. Errors and uncertainties observed in this phase depend mainly on the latency to move the packet from the network interface to the kernel and then from the kernel to the user space. Furthermore, the OS scheduler could assign the computing resources to another concurrent process before the second packet arrives at the receiver. By adopting the same approach to that found in the packet-sending phase, we estimated the latency $\Delta t$ to move the packet from the kernel to the user space. This latency varies from 5 to 65 microseconds.

Spruce, IGI and Pathchirp are free from errors and uncertainties due to received packets timestamping. Indeed, these tools deal with errors by timestamping packets directly in the kernel space. To do this, Spruce and Pathchirp use SO_TIMESTAMP option in a SOCK_DGRAM socket that enables *recvmsg()*-call to return a timestamp corresponding to the time the packet was received at the socket level. To grab received packets from the network interface and to timestamp them directly, IGI uses the libpcap module, a packet capture library that provides implementation-dependent access to the underlying packet capture facilities provided by the operating system. Delays and jitters measured by Pathload are large enough to make the error due to timestamping operations negligible.

The tools based on the Probe Gap Model suppose that the path capacity is known and use this metrics to estimate the available bandwidth. Indeed, Spruce and IGI separate capacity measurement from available bandwidth measurement and Spruce assumes that this metric is known and keeps stable when measuring available bandwidth. IGI assumes that capacity can be easily measured with one of the capacity measurement tools (Nettimer [8], Pathrate [13], Bprobe [9] etc). Since the available bandwidth is obtained by subtracting the estimated cross traffic throughput from an estimate of the bottleneck link capacity, an error in capacity estimation will directly impact the accuracy of available bandwidth measurements.

We apply the differential calculus method in order to determine the available bandwidth measurement uncertainty obtained using IGI. According to formula (4) given in section 2, errors on IGI are estimated as:

$$\Delta A = \left| 1 - \frac{D_{out}}{D_{in}} \right| . \Delta C \tag{9}$$

Assuming that the path capacity is *C=97.5 Mbs* (IP layer), an error of *ΔC=20Mbs* (about *20%* of *C*) will involve an uncertainty of *ΔA=11.6 Mbs* (around *29%* of *A*) when the available bandwidth *A=40 Mbs*.

In IGI measurement, supplementary errors on available bandwidth measurement are experienced due to the packet correlation phenomenon. Indeed, IGI uses periodic probe packet streams and in a given stream, three successive packets *k, k +1* and *k +2* constitute two successive packet-pairs $P_1$=(k , k+1) and $P_2$=(k+1 , k+2). Since each packet-pair depends on the next, the inter-packet gaps measured are correlated. To minimize the impact of this behaviour on the available bandwidth measurement, IGI does not take into account packets that have final inter-packet gap $D_{out}$ less or equal to the initial inter-packet gap $D_{in}$ but considers them as cross traffic packets. This phenomenon leads to an overestimation of the cross traffic throughput and, as a result, to an underestimation of the available bandwidth measurement.

## 6   Conclusion

In this article we have presented an analysis and a comparative study of available bandwidth measurement techniques and we have evaluated different measurement tools on similar testbed configurations. We have compared tools performance in terms of accuracy, intrusiveness and response time. Furthermore, we have tried to analyse and to explain errors and uncertainties observed on the tools we have studied. The results obtained show that Pathload is the most intrusive tool and, in some cases, can be very slow. Pathchirp hugely overestimates the available bandwidth and IGI is inaccurate. Finally, Spruce seems to be the tool that offers the best performance with regards to the criteria studied. It is the fastest and most accurate tool and one of the least intrusive.

The study presented in this paper was focused on the small number of criteria that seemed to be the most important. However, this study must be completed by considering other parameters and by evaluating the measurement tools on real networks settings.